# Machine-Learning Enabled Multidimensional Data Utilization Through Multi-Resonance Architecture: A Pathway to Enhanced Accuracy in Biosensing

Majid Aalizadeh[1,2,3,4], Morteza Azmoudeh Afshar[5], and Xudong Fan[1,3,4,*]

[1]Department of Biomedical Engineering,
University of Michigan, Ann Arbor, MI 48109, USA

[2]Department of Electrical Engineering and Computer Science,
University of Michigan, Ann Arbor, MI 48109, USA

[3]Center for Wireless Integrated MicroSensing and Systems (WIMS[2]),
University of Michigan, Ann Arbor, MI 48109, USA

[4]Max Harry Weil Institute for Critical Care Research and Innovation,
University of Michigan, Ann Arbor, MI 48109, USA

[5]Informatics Institute
Istanbul Technical University, 34485 Istanbul, Turkey

*: Corresponding author: xsfan@umich.edu

**Abstract**

A novel framework is proposed that combines multi-resonance biosensors with machine learning (ML) to significantly enhance the accuracy of parameter prediction in biosensing. Unlike traditional single-resonance systems, which are limited to one-dimensional datasets, this approach leverages multi-dimensional data generated by a custom-designed nanostructure—a periodic array of silicon nanorods with a triangular cross-section over an aluminum reflector. High bulk sensitivity values are achieved for this multi-resonant structure, with certain resonant peaks reaching up to 1706 nm/RIU. The field analysis reveals Mie resonances as the physical reason behind the peaks. The predictive power of multiple resonant peaks from transverse magnetic (TM) and transverse electric (TE) polarizations is evaluated using Ridge Regression modeling. Systematic analysis reveals that incorporating multiple resonances yields up to three orders of magnitude improvement in refractive index detection precision compared to single-peak analyses. This precision enhancement is achieved without modifications to the biosensor hardware, highlighting the potential of data-centric strategies in biosensing. The findings establish a new paradigm in biosensing, demonstrating that the synergy between multi-resonance data acquisition and ML-based analysis can significantly enhance detection accuracy. This study provides a scalable pathway for advancing high-precision biosensing technologies.

## 1. Introduction

A notable innovation in optical biosensing is the use of resonant structures that enhance detection performance through light-nanostructure interactions [1-9]. Traditionally, single-resonance biosensors offer a single measurable response to external changes like refractive index variations [10-16]. Refractive index sensors are used in biosensing and medical diagnostics, chemical and environmental sensing, food and beverage quality control, optical and photonic devices, pharmaceutical analysis, industrial monitoring [17-24]. Index However, the limitations of single-resonance systems [25, 26] in capturing complex, multi-dimensional biological signals have led to the exploration of new approaches. Introducing multi-resonant structures [27-31] presents a transformative advancement in the field, enabling the collection of richer, multi-dimensional datasets and paving the way for improved precision and robustness in parameter prediction.

Machine learning (ML) has become a groundbreaking tool in biosensing, facilitating the analysis and interpretation of complex, high-dimensional data [32-36]. Combining multi-resonance biosensors with ML methods can significantly enhance the precision of target variable predictions. This combination is particularly effective with multi-dimensional datasets that contain latent correlations and patterns not evident in lower-dimensional data. Traditional one-dimensional (1D) biosensing data have often been modeled using simple linear fittings, which may not provide optimal precision [37, 38].

Here, we propose a novel framework to overcome these limitations by utilizing multi-dimensional data and demonstrate the high potential of utilizing multi-dimensional data in biosensing through an exemplary optical multi-resonant structure. Our approach is based on the principle that incorporating multiple predictors or resonances leads to superior prediction accuracy of the target variable compared to using lower-dimensional or 1D data. This hypothesis is validated by modeling a relatively simple-to-fabricate nanostructure consisting of a triangular cross-sectioned array of periodic silicon nanorods over an aluminum back reflector.

Most resonant peaks in our structure exhibit sensitivity values in the hundreds of nm/RIU, with specific peaks reaching bulk sensitivity as high as 1706 nm/RIU. It is shown through electric and magnetic field analysis that the magnetic and electric dipole type Mie resonances are the physical reasons behind the TM and TE peaks, respectively. Such sensitivity performance underscores the effectiveness of our design [39] and exemplifies the broader principle demonstrated in this work, that is, utilizing multi-dimensional data can significantly enhance biosensing precision. To quantitatively establish this principle, we select four resonant peaks from the absorption profiles of transverse magnetic (TM) and transverse electric (TE) polarizations and employ ML techniques, namely the Ridge Regression modeling, to analyze their predictive power [40-44]. By systematically evaluating all possible combinations of these predictors (peaks)—from using each individually to incorporating all four simultaneously—we show that adding predictors consistently improves precision. Specifically, when combining the peaks from both polarizations, approximately 3 orders of magnitude enhancement in bulk refractive index detection precision is achieved compared to some single peaks. This approach allows us to maintain the same hardware—biosensor structure and measurement setup—while enhancing performance through innovative data processing. The findings presented in this paper not only highlight the desirable sensitivity performance of our multi-resonant structure but also provide proof of concept for a general novel paradigm in biosensing. By demonstrating the synergy between multi-resonance data acquisition and ML-based analysis, this work lays the foundation for future advancements in high-precision biosensing technologies.

## 2. Proposed structure

The schematic of the proposed structure is shown in Fig. 1 along with the definition of the xyz coordinates. The structure consists of a periodic array of silicon based nanorods with triangular cross-section. Figure 1(b) demonstrates the cross-section (side view) of the structure, where it can be observed that the nanorods have a triangular shape in the cross section with the base (or array period), p = 2 μm, and height, h = 4 μm. The reason to use a tapered design instead of one with vertical sidewalls is that a tapered design is equivalent to superposition of various widths, each supporting a specific resonance, leading to maximization of the multi-resonant behavior. The substrate is an optically thick Aluminum layer to ensure that all the transmission is blocked.

## 3. Simulation results

The reflection spectrum of the proposed structure is simulated. The simulations are carried out in the Finite-difference time-domain (FDTD) environment using the commercial Ansys Lumerical FDTD software, and the details of the simulation are mentioned in the "Methods" section. Since the transmission (T) is almost zero, the absorption spectrum (A) can be directly found from the reflection spectrum (R), *i.e.*, A=1-R, which is exemplified in Figs. 2(a) and (b) for the spectral range of 0.4-7 μm for the TM and TE polarizations, respectively. It can be seen that for both polarizations there are a high number of distinct peaks in the absorption spectrum, covering the infrared range. There are also a high number of peaks available in the wavelength range below 1 μm that cannot be seen very distinctly in Figs. 2(a,b). Therefore, Figs. 2(c,d) show the TM and TE absorption spectra of the structure in the 0.4-1 μm spectral range to demonstrate the peaks in that range more clearly. However, for simplicity, in the rest of this work, the peaks in the infrared range are used to demonstrate the concept.

To evaluate the biosensing functionality of the structure, we examine the peak shift with varying bulk refractive index above the structure. Figure 3(a) shows the absorption spectra of the TM polarization for the three bulk refractive index values of 1.45, 1.50, and 1.55. It can be observed that the peaks undergo a red shift (towards higher wavelengths) as the bulk refractive index increases. For example, Fig. 3(b) shows the close-up view of the shift of the absorption peak around the wavelength of 3.3 μm. It can be observed that the peak shifts around 139.2 nm for a 0.1 increment in the bulk index (from 1.45 to 1.55), corresponding to a bulk sensitivity of 1392 nm/RIU for this peak. Another example of a TM polarization peak with high sensitivity is the one around the wavelength of 4.8 μm which has a bulk sensitivity of 1149 nm/RIU, shown in Fig. 3(c) (See Table 1 for more peak sensitivity examples for both polarizations).

Figure 3(d) shows the absorption spectra shift of the TE polarization for the bulk refractive index values of 1.45, 1.50, and 1.55. Fig. 3(e) shows the close-up view of the shift of the absorption peak around the wavelength of 3.5 μm, which has a bulk sensitivity of 1283 nm/RIU. Another example is the peak around the wavelength of 4.2 μm. The bulk sensitivity is calculated to be around 784 nm/RIU for this peak. Unlike other designs requiring intricate multilayered nanofabrication, our design achieves a high sensitivity performance with enhanced simplicity and scalability.

By comparing the peak shifts for both polarizations, we can see that almost all the peaks for the TM polarization undergo a stronger red-shift compared to the TE polarization. This is due to the fact that the electric field is parallel to the cross-section plane for the TM polarization, and hence it penetrates directly into the surface of the Si triangular nanorods (see Fig. 1(b)). Therefore, the electric resonances for the TM polarization can be influenced more significantly by the index variation above the surface of the structure. This is while for the TE polarization, the electric field is perpendicular to the cross-section, *i.e.*, parallel to the surface of triangular nanorods, making the electric resonances less sensitive to index variations.

To evaluate the refractive index sensitivity of the structure more closely, we vary the refractive index of the simulation environment, from 1.45 to 1.55 by 0.001 increments. For the TM and TE polarizations, certain peaks shown in Figs. 3(a) and 3(d) are chosen, respectively, and the wavelength shift versus the bulk index shift are plotted for those peaks. The results are shown in Figs. 4(a,b) for the TM and TE polarizations, respectively. The nm/RIU for the peak shifts in Fig. 4 are presented in Table 1.

As seen in Fig. 4, like most biosensing measurements, the relationship between the measured quantity (peak shift) and the sensing input (bulk index) seems linear. However, even in the simulations, the relationships are not perfectly linear due to nonlinear physical phenomena such as adjacent resonances or peaks influencing each other. In actual experimental measurements, the linearity may be even weaker due to experimental imperfections and errors such as noise or measurement setup inaccuracies. Therefore, a purely linear model does not always provide a precise prediction of the bulk index value based on the measured resonance location. This occurs when only one peak shift is considered, and a linear fit is used as the calibration curve. However, as will be demonstrated in detail in the coming sections, if, through data training and machine learning algorithms, more than one peak (in this case 4 peaks) is considered, the precision of estimating the bulk index based on the measured peak location gets significantly enhanced. In other words, if we use multiple peaks or resonances, or a multi-dimensional set of data (4D in this case), instead of one peak, or using 1D data, the precision of the system will be highly improved. This enables us to attain significantly higher precision using the same biosensor structure and the same measurement setup, but with a slightly different post data processing strategy.

Figures 5 and 6 present the electric and magnetic field distributions for the selected TM and TE polarization resonances, respectively, providing strong confirmation of Mie-type dipole modes within the silicon nanorod [45]. For TM polarization (Figure 5), where the incident wave consists of $E_x$ and $H_z$, the fields inside the structure include $E_x$, $E_y$, and $H_z$. The bottom row of Figure 5 illustrates the localized Hz hotspots, while the top row presents quiver plots of $E_x$-$E_y$, revealing clear circulating electric field loops around the magnetic field peaks. This behavior aligns with Ampère's Law, $\nabla\times E=-\partial B/\partial t$, where a time-varying $H_z$ dipole moment along the z-axis induces a rotational $E_x$, $E_y$ field around it. The presence of these closed-loop electric field structures around a concentrated magnetic field hotspot confirms that these resonances correspond to magnetic dipole Mie modes.

Similarly, Fig. 6 (TE polarization) shows the field behavior for an incident wave consisting of $H_x$ and $E_z$, with the fields inside the nanorod including $H_x$, $H_y$, and $E_z$. The bottom row illustrates the localized $E_z$ field hotspots, while the top row quiver maps display circulating $H_x$, $H_y$ vectors around these regions. This pattern is characteristic of Faraday's Law, $\nabla \times H = \partial D / \partial t$ , where an oscillating electric dipole moment along the z-axis ($E_z$) induces a surrounding rotational magnetic field ($H_x$, $H_y$). The formation of these looped H-field structures around strong $E_z$ hotspots confirms that these TE resonances correspond to electric dipole Mie modes.

Together, these field distributions validate that TM polarization supports magnetic dipole Mie resonances, while TE polarization supports electric dipole Mie resonances, fully consistent with the expected behavior of high-index dielectric nanostructures in the infrared. The presence of well-defined dipole fields and circulating response patterns further reinforce that these resonances originate from localized Mie-type scattering rather than extended traveling waves, governing the observed refractive index sensitivity trends.

## 4. Multi-resonance data processing through ML and precision enhancement

To demonstrate the biosensing precision enhancement through ML based multi-dimensional data processing, we consider the shift of 4 TM and 4 TE peaks shown in Figs. 4(a) and 4(b), respectively. Figures 7(a) and 7(b) demonstrate the scatter plots of the shift of those 4 peaks for TM and TE polarizations, respectively, versus the bulk index shift starting from 1.45. The bulk index shift ranging between 0.001 to 0.1 with 0.001 increments is multiplied by 1000 for simplicity, and is denoted as x, *i.e.*, 100 values in total, ranging from 1 to 100. The peak shift values are referred to as $y_1$, $y_2$, $y_3$, and $y_4$ for the TM peaks at the wavelengths of 1859, 2926, 3281, and 4854 nm, respectively, and as $y_{11}$, $y_{22}$, $y_{33}$, and $y_{44}$ for the TE peaks at the wavelengths of 2349, 3428, 3744, and 4189 nm, respectively. Since y values (peak shifts) are measured in the biosensor (the location of peaks) and are used to predict the value of x associated with them (bulk index shift), $y_i$'s are referred to as predictors, and x is referred to as the target variable. Table S1 in the Supplementary Information shows the descriptive statistics of the target variable x and the TM polarization predictors. The model development for ML based data training is detailed in the Supplementary Information, using the TM polarization predictors for demonstration. During model development, we conducted multicollinearity assessments and calculated Variance Inflation Factors (VIF) to ensure predictor robustness. VIF values suggested strong multicollinearity among predictors. Therefore, in addition to the linear regression used for multi-dimensional modeling initially, we have used a Ridge Regression model to address possible adverse effects of multicollinearity in multi-dimensional data modeling. Additionally, all models underwent rigorous validation using 10-fold cross-validation to ensure reliability and generalizability.

Figure 8 illustrates the performance enhancement achieved by leveraging multiple resonance shifts in the proposed multi-resonance optical biosensor for bulk refractive index sensing. The diagrams in Figs. 8(a) and (b) are organized to depict all possible combinations of predictors $y_1$, $y_2$, $y_3$, and $y_4$, and their associated mean squared error (MSE) values using the Ridge Regression modeling for the TM and TE polarizations, respectively. The leftmost column represents the individual predictors (‘$y_1$’ through ‘$y_4$’) and their corresponding MSE when used individually in simple 1D linear modeling. As we move rightward, the combinations of the predictors increase, culminating in the simultaneous utilization of all four predictors. Each box is labeled with its respective MSE value, showcasing the performance of that specific combination of predictors.

The progression from left to right consistently demonstrates a reduction in MSE values as additional predictors are incorporated. For instance, in the TM polarization, two peaks used individually yield MSE values of 2.466 and 7.226, respectively. However, when all four TM peaks are utilized together, the MSE dramatically reduces to 0.0173. Similarly, in the TE polarization, individual MSEs of 1.1186 and 7.3318 are reduced to 0.0357 when all four TE peaks are combined. Importantly, by incorporating both TM and TE polarizations and utilizing all eight resonant peaks as 8-dimensional data, the MSE further improves to 0.0090, with an $R^2$ value of 1.000, demonstrating near-perfect precision.

The lines between the columns connect the combinations in the right columns that can be produced by their connected adjacent left columns by adding another predictor to it. There are 28 connecting lines in total. The red line highlights combinations where the inclusion of additional predictors results in a two-order-of-magnitude improvement in MSE. For instance, for TM polarization, combining ‘$y_2$’ and ‘$y_3$’ leads to a 274-fold precision gain compared to using ‘$y_2$’ alone. There are in total 9 blue lines which indicate at least one order of magnitude improvement, while black lines represent MSE enhancements ranging between 1-10 times. This means for a total of 10 out of 28 lines, adding an additional predictor to certain predictor combinations, leads to at least an order of magnitude enhancement in the modeling precision for the case of TM polarization, which is over 35% of all the lines. For the case of TE polarization, there are 2 red lines (>100-fold enhancement) and 8 blue lines (>10-fold enhancement), indicating a similar precision enhancement performance compared to the TM polarization. The yellow lines represent combinations where the inclusion of additional predictors results in no improvement. There is only one yellow line for each of the TM and TE polarizations in Figs. 8(a,b), out of a total of 28 cases, where in both cases MSE has remained approximately the same. It can therefore be observed that additional predictors have nonlinear impact. While the overall trend is a reduction in MSE with more predictors, the magnitude of improvement varies. For example, the addition of '$y_4$' to '$y_1$' and '$y_3$' (indicated by black lines) provides less significant improvement compared to the addition of '$y_2$' to '$y_1$' (red line) for the TM polarization.

The findings illustrated by the figure confirm that multi-resonance analysis significantly enhances biosensing precision without requiring additional hardware. This improvement stems solely from advanced post-processing using machine learning. The systematic reduction in MSE demonstrates the ability of machine learning techniques to extract synergistic information from multiple resonance shifts, effectively overcoming limitations of traditional single-resonance methods.

Figure 8, as a complete picture, further emphasizes the cost-effectiveness of the proposed methodology. Traditional methods rely on hardware advancements to improve sensing precision, often leading to increased costs and complexity. In contrast, this approach achieves superior performance solely through data-driven techniques applied to existing hardware. This paradigm shift not only reduces costs but also broadens the applicability of high-precision biosensors in resource-constrained settings.

Finally, combining all the peaks in both TE and TM polarizations, *i.e.*, using 8 dimensional data in our Ridge Regression model, the MSE and $R^2$ values of 0.009 and 1.000 are achieved, respectively. This further attests to the enhancement of accuracy in data modeling when all the predictors are incorporated. Other parameters associated with the 8-dimensional modeling are available in the Supplementary Information.

## 5. Discussion, conclusion, and future work

This study establishes a significant advancement in the field of biosensing by demonstrating the transformational potential of utilizing multidimensional data in multi-resonance biosensors. The results provide compelling evidence that the integration of multiple predictors significantly enhances the precision of target variable prediction. By using a simple yet highly effective optical multi-resonant structure as an illustrative example, we show that the principle of multidimensional data utilization is not only theoretically sound but also practically achievable with scalable fabrication techniques such as nanoimprint lithography.

Our structure, consisting of a triangular cross-section array of periodic silicon nanorods over an aluminum back reflector, exemplifies a design that combines simplicity with high performance. The sensitivity of the structure, which can be as high as 1706 nm/RIU, underscores its capability as a biosensing platform. This performance is further elevated when the multi-dimensional nature of its resonances is leveraged through machine learning methods. By systematically analyzing combinations of four resonant predictors using Ridge Regression modeling, we reveal a clear trend: the inclusion of additional predictors leads to consistent improvements in precision, with some combinations achieving over an order of magnitude reduction in mean squared error.

Notably, for TM polarization, individual peaks result in MSEs of 2.466 and 7.226, while the use of all four peaks reduces the MSE to 0.0173. Similarly, for TE polarization, the individual MSEs of 1.1186 and 7.3318 are reduced to 0.0357 when all four peaks are combined. Furthermore, combining TM and TE polarizations and utilizing all eight resonant peaks results in the best performance observed in this study, with an MSE of 0.0090 and an $R^2$ value of 1.000. This highlights the high potential of maintaining the same biosensor structure and measurement setup while enhancing performance through advanced data processing techniques. This approach represents a novel and efficient pathway to significantly improve biosensing precision without the need for complex hardware modifications.

The ability to achieve near-perfect precision with a simple biosensor design and advanced data processing renders this work particularly promising for complex and real-time biosensing applications. The significant enhancement in data interpretation accuracy demonstrated here is critical for addressing the challenges of high-speed and high-complexity biosensing environments. By providing a robust framework for leveraging multidimensional data, this work paves the way for biosensing platforms capable of tackling intricate biological systems and dynamic scenarios.

The implications of these findings extend beyond the specific design and application studied here. They represent a general principle that can be applied to a wide range of biosensing technologies. Multidimensional data acquisition, when coupled with advanced data analysis techniques such as machine learning, offers a powerful pathway to overcoming the limitations of traditional one-dimensional biosensing approaches. This paradigm shift has the potential to significantly enhance the accuracy, reliability, and robustness of biosensors across diverse applications.

Looking ahead, the integration of multi-resonance biosensors with sophisticated ML algorithms opens new avenues for innovation in biosensing. Future research could explore the application of this principle to more complex biological systems, incorporate additional predictors from other physical modalities, and develop advanced machine learning frameworks tailored for biosensing data. By advancing the understanding and application of multidimensional data utilization, this work contributes to the ongoing evolution of biosensing technologies and their impact on science and society.

**6. Methods**

Lumerical simulation setup: The simulation is carried out in the FDTD simulation environment using the Ansys Lumerical commercial software. The simulation environment is 2D in the xy plane since the structure is infinite in the z axis direction. In other words, only the cross section of one period of the structure is being simulated. The boundary condition in the x axis direction is the periodic boundary condition, while along the y axis it is chosen to be the perfectly matched layer (PML) boundary condition to ensure no reflection back in that direction. The Frequency-Domain Field Profile DFT monitor is placed right under the top boundary of the FDTD simulation environment, and the planewave source is placed under the monitor. The monitor collects the reflection data. The monitor has 30000 frequency points and collects the spectrum in the 0.4-10 μm range. The planewave source is incident in the -y direction for the normal incidence.

For the field simulations, the monitor was placed in the xy plane in the simulation environment. The field simulations are carried out with the 1.45 bulk index since the resonant behavior of certain peaks may disappear while changing the bulk index from 1 to 1.45.

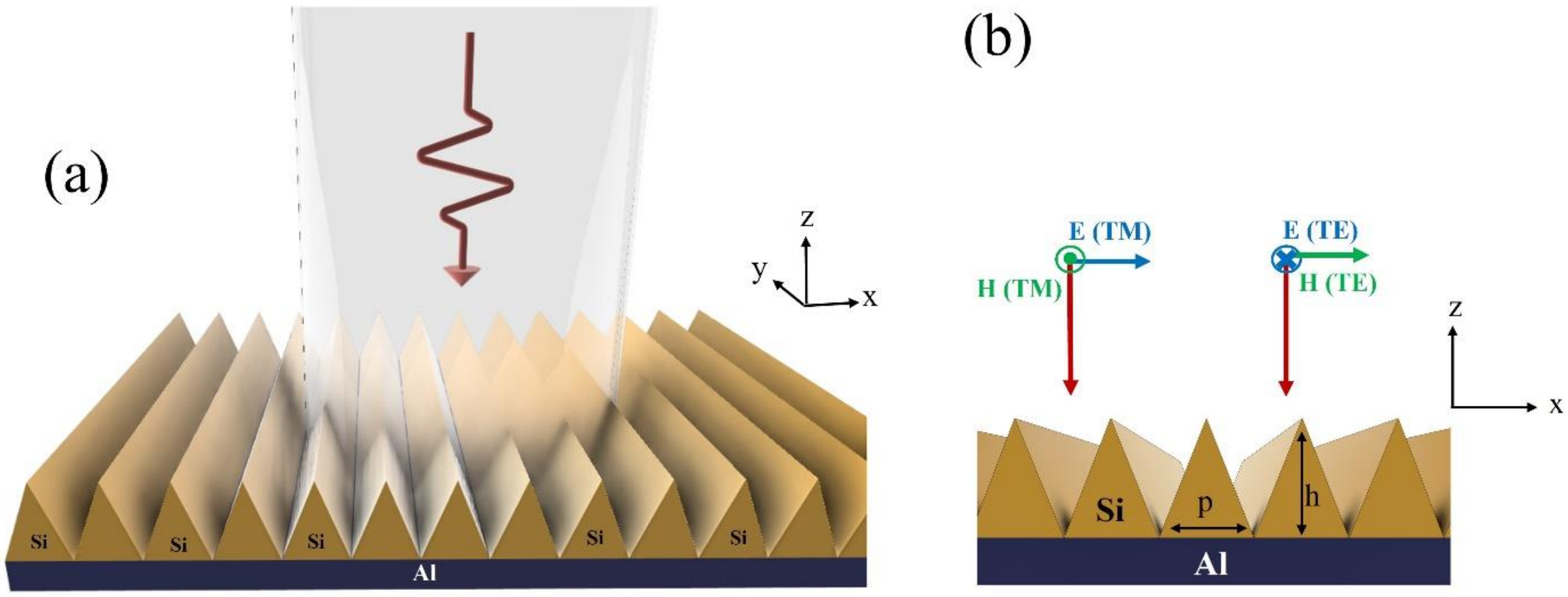

Figure 1. (a) 3D view and (b) cross-section view of the proposed design, consisting of a periodic array of triangular cross sectioned Si nanorods over an Al back reflector. p and h denote the period and the height of nanorods, respectively. p = 2 μm and h = 4 μm. The direction of the electric (in blue) and magnetic (in green) fields for the TM and TE polarizations of the incident beam are also shown.

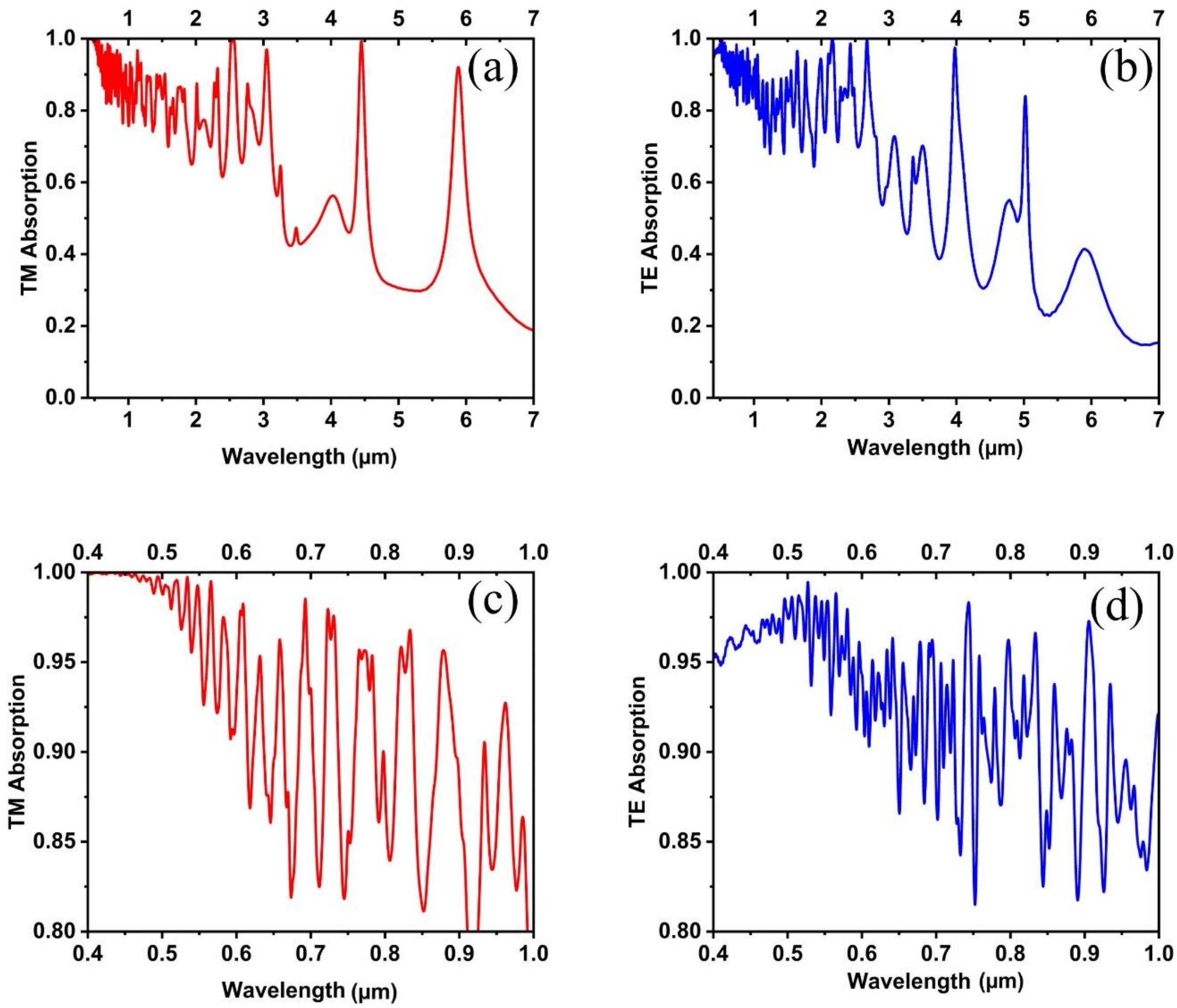

Figure 2. Absorption spectra for (a,c) TM and (b,d) TE polarizations at two different spectral ranges.

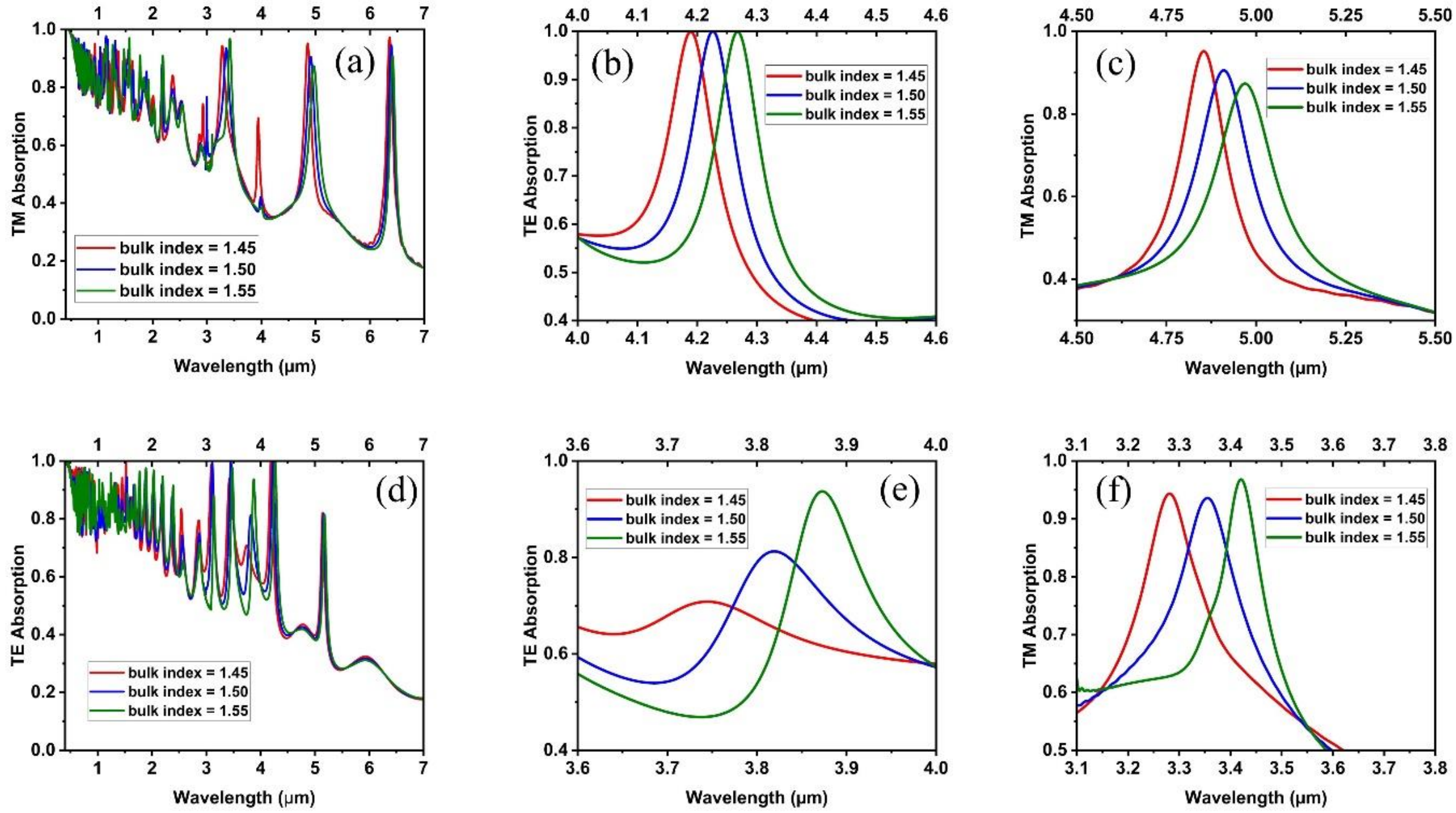

Figure 3. Absorption spectra for three different bulk refractive index values of 1.45, 1.50, and 1.55 for the (a-c) TM, and (d-f) TE polarizations.

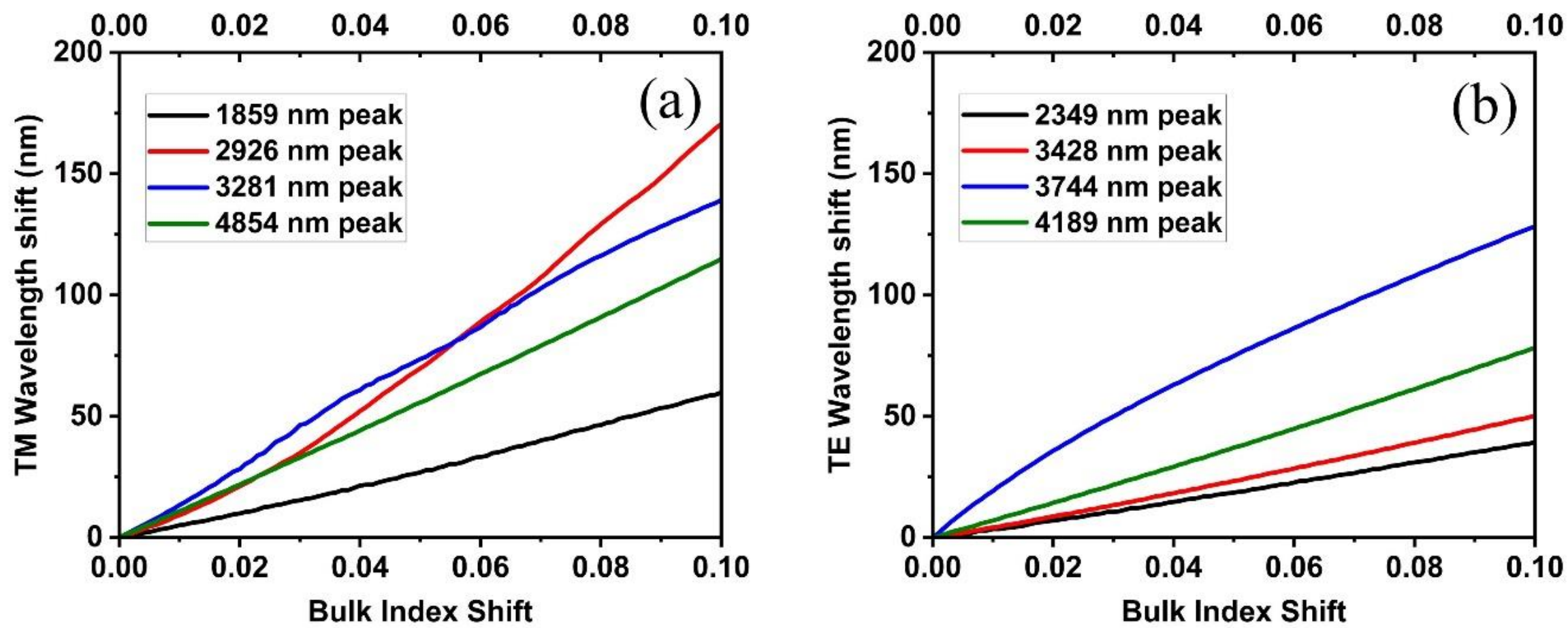

Figure 4. Resonance wavelength shift vs bulk index change for the structure for 4 peaks of (a) TM and (b) TE polarization. The peak locations (in nm) mentioned in the figure refer to their location at the bulk index of 1.45.

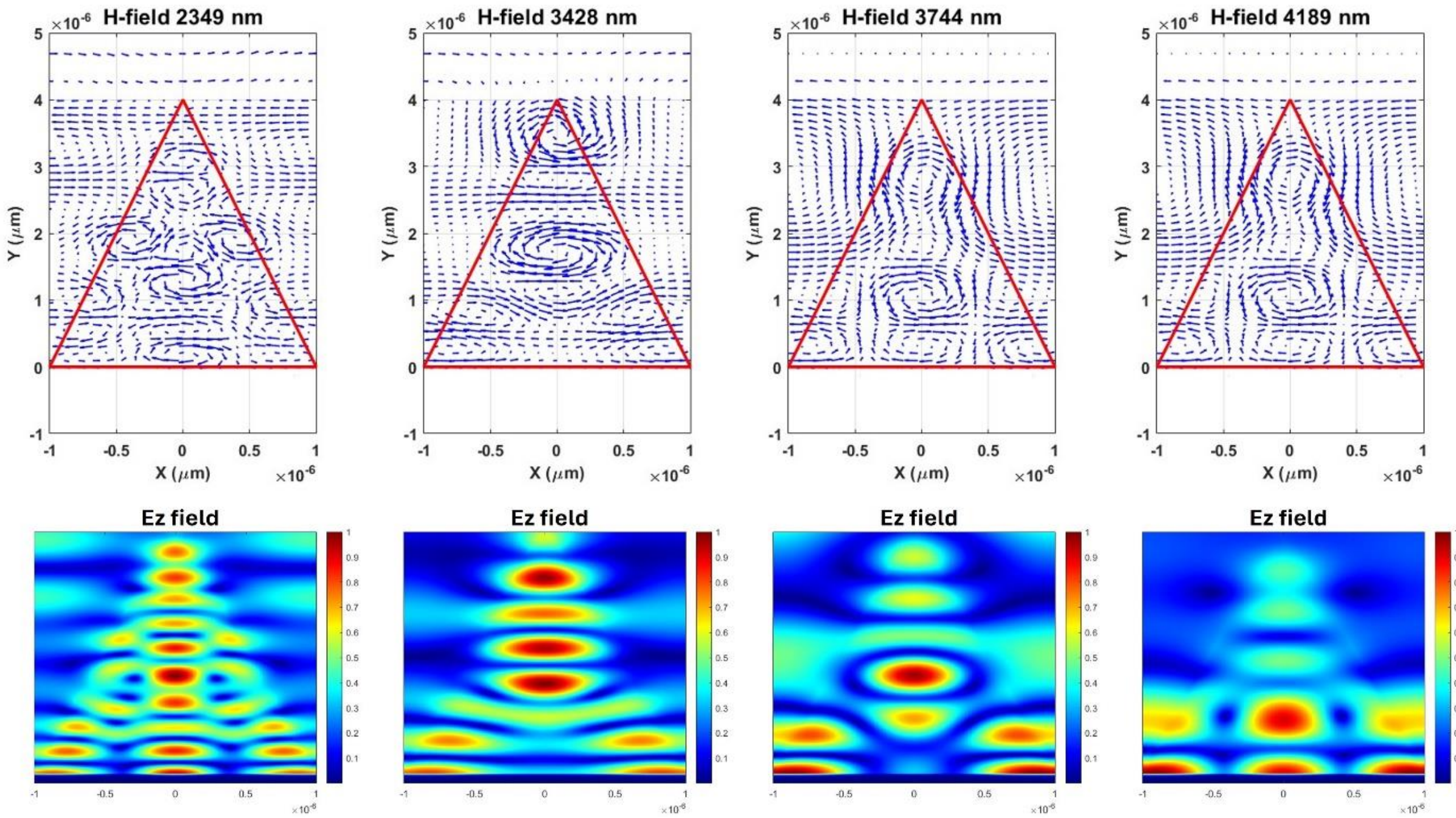

Figure 5. H-field quiver maps (top panel) and Ez field magnitude color maps (bottom panel) of the TM 4 peaks selected for analysis listed in Table 1.

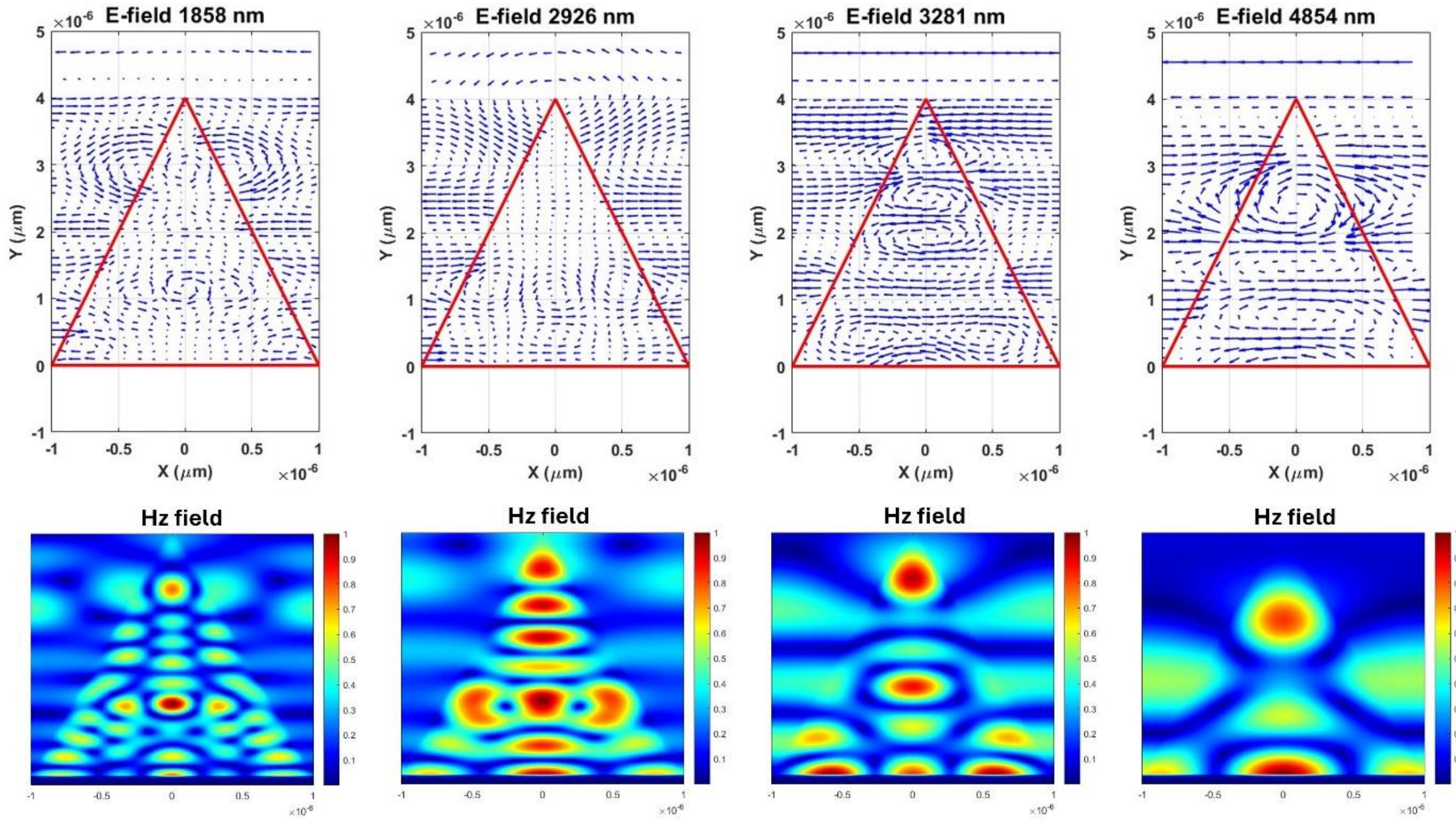

Figure 6. E-field quiver maps (top panel) and Hz field magnitude color maps (bottom panel) of the TE 4 peaks selected for analysis listed in Table 1.

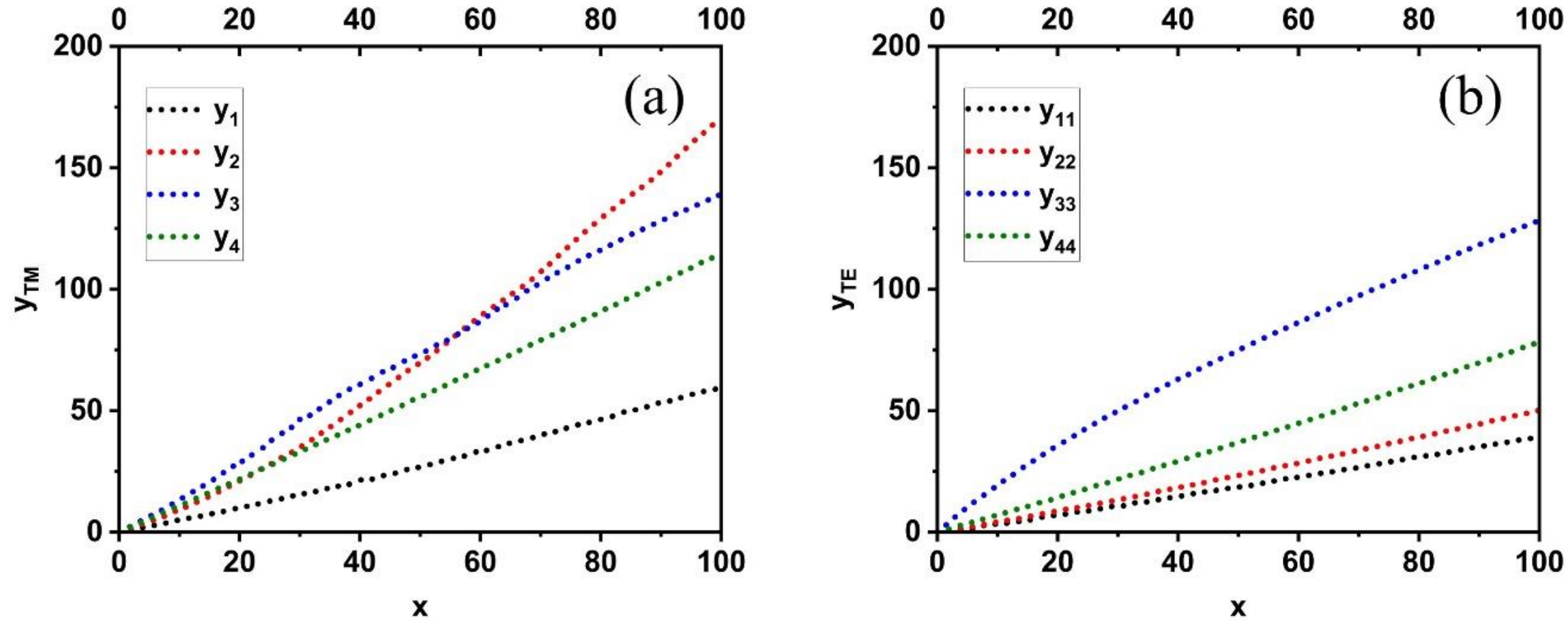

Figure 7. Scatter plots of $y_i$'s vs x, referring to the peaks chosen from the (a) TM and (b) TE absorption spectra vs bulk index change (1000x).

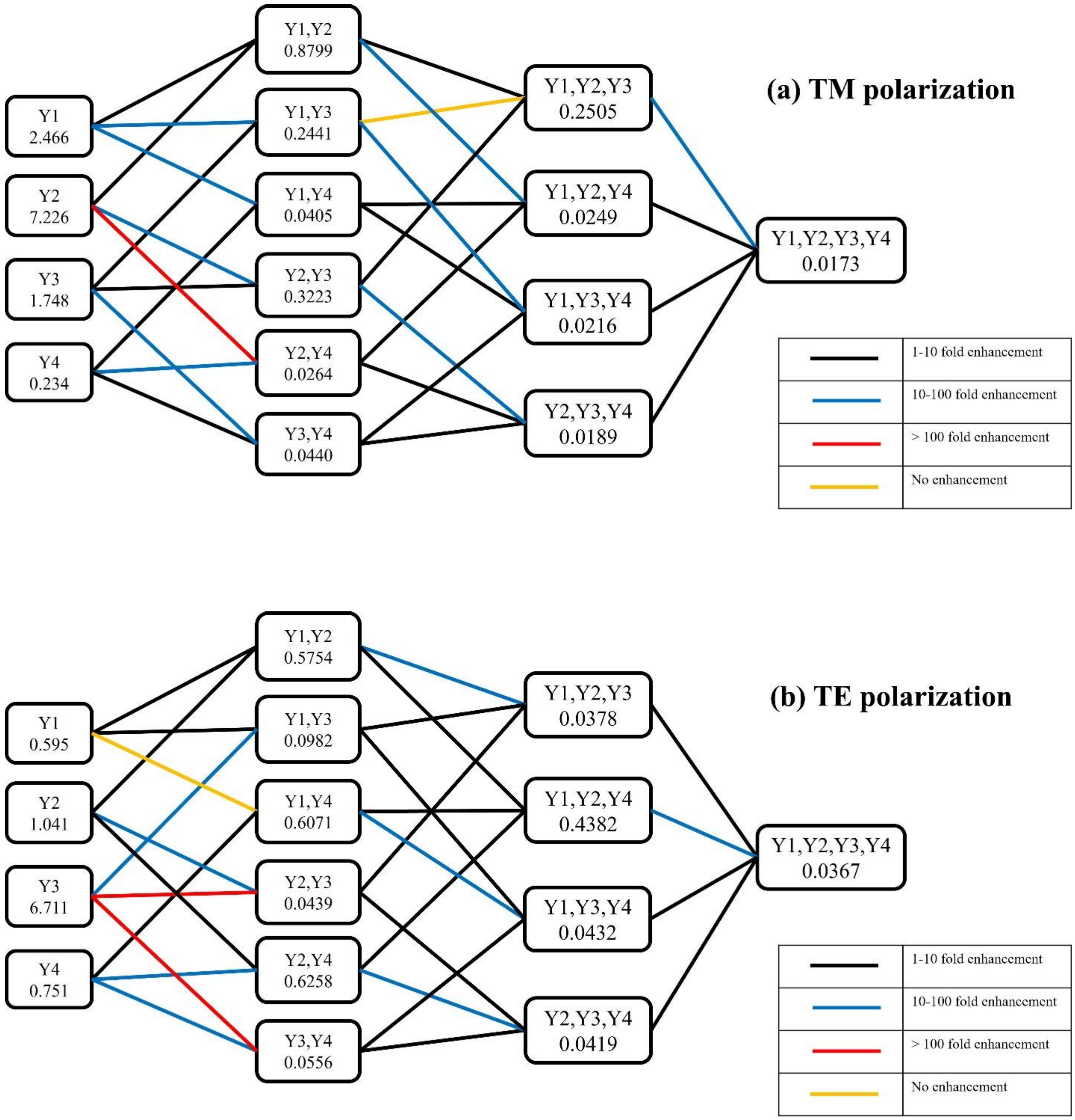

Figure 8. MSE diagram comparing the accuracy performance of Ridge Regression modeling using all possible combinations of the four predictors for (a) TM and (b) TE polarizations.

Table 1. nm/RIU for the peak shifts in Fig. 4.

| TM Polarization | | TE Polarization | |
|---|---|---|---|
| Peak Location (nm)[1] | Sensitivity (nm/RIU) | Peak Location (nm) | Sensitivity (nm/RIU) |
| 1859 | 598 | 2349 | 394 |
| 2926 | 1706 | 3428 | 502 |
| 3281 | 1392 | 3744 | 1283 |
| 4854 | 1149 | 4189 | 784 |

Note 1: The peak locations (in nm) mentioned in the table refer to their locations at the bulk index of 1.45.

# Supplementary Information for

# Machine-Learning Enabled Multidimensional Data Utilization Through Multi-Resonance Architecture: A Pathway to Enhanced Accuracy in Biosensing

Majid Aalizadeh[1,2,3,4], Morteza Azmoudeh Afshar[5], and Xudong Fan[1,3,4,*]

[1]Department of Biomedical Engineering,
University of Michigan, Ann Arbor, MI 48109, USA

[2]Department of Electrical Engineering and Computer Science,
University of Michigan, Ann Arbor, MI 48109, USA

[3]Center for Wireless Integrated MicroSensing and Systems (WIMS$^2$),
University of Michigan, Ann Arbor, MI 48109, USA

[4]Max Harry Weil Institute for Critical Care Research and Innovation,
University of Michigan, Ann Arbor, MI 48109, USA

[5]Informatics Institute
Istanbul Technical University, 34485 Istanbul, Turkey

address

*: Corresponding author: xsfan@umich.edu

**S1. Descriptive statistics of all variables for the TM polarization**

Table S1. Descriptive statistics of variables for the TM polarization

| Statistic | x | $y_1$ | $y_2$ | $y_3$ | $y_4$ |
|---|---|---|---|---|---|
| Count | 100 | 100 | 100 | 100 | 100 |
| Mean | 50.5000 | 28.4201 | 75.4105 | 73.3240 | 56.8787 |
| Std. Deviation | 29.0115 | 17.6078 | 51.3501 | 41.3989 | 33.3703 |
| Minimum | 1.0000 | 0.6400 | 0.9600 | 1.2800 | 1.2800 |
| 25th Percentile | 25.7500 | 13.2804 | 28.8010 | 38.8013 | 28.2409 |
| Median | 50.5000 | 27.3609 | 70.5624 | 74.0825 | 56.1619 |
| 75th Percentile | 75.2500 | 43.6815 | 118.9640 | 110.1637 | 85.1228 |
| Maximum | 100.0000 | 59.8420 | 170.5657 | 139.2046 | 114.8838 |

**S2. Correlation analysis**

The Pearson correlation coefficients, presented in Table S2, show extremely strong positive correlations between x and each $y_i$, and between the $y_i$'s themselves. The exploratory analysis indicates that while each predictor is strongly related to x, the high correlations among predictors suggest multicollinearity, which can affect multiple regression models. This necessitates careful modeling strategies to mitigate multicollinearity's impact.

Table S2. Correlation matrix.

| | x | $y_1$ | $y_2$ | $y_3$ | $y_4$ |
|---|---|---|---|---|---|
| x | 1.0000 | 0.9986 | 0.9958 | 0.9990 | 0.9999 |
| $y_1$ | 0.9986 | 1.0000 | 0.9991 | 0.9957 | 0.9993 |
| $y_2$ | 0.9958 | 0.9991 | 1.0000 | 0.9915 | 0.9971 |
| $y_3$ | 0.9990 | 0.9957 | 0.9915 | 1.0000 | 0.9982 |
| $y_4$ | 0.9999 | 0.9993 | 0.9971 | 0.9982 | 1.0000 |

The plots in Fig. 5 indicate strong positive linear relationships, suggesting that linear regression models may be appropriate.

**S3. 10-fold cross validation**

It is noteworthy that the number of observations data is 100 in this case. In other words, the data can be 1D, meaning 100 $(x,y_i)$ pairs for each individual $y_i$, or can be 4D, meaning 100 sets of $(x, y_1,y_2,y_3,y_4)$. To develop the models and assess their accuracy, a 10-fold cross-validation approach is carried out for all models. In general, a k-fold cross validation is a technique which works in the following way: First, the data is shuffled and put it in k equal groups or folds. Then on each iteration, the remaining (k-1) set of folds (or groups) are used as our training set and the metrics such as accuracy are measured in the test set, which is the remaining 1 group (or fold) out of the k total groups. This process is repeated in each iteration, until all the k sets are used once as a test set (and the other k-1 groups as training sets) to measure the metrics. The final metric such as accuracy is the average of all the accuracies from each iteration. This method is used when the number of observations is relatively small (100 in this case), and the option to create a large test set is not available. Therefore, all models are developed using 10-fold cross-validation, where the dataset is divided into 10 folds, and the model is trained and tested 10 times, each time using a different fold as the test set.

**S4. Simple 1D linear regression models**

First to test the precision of linear fitting for 1D data, for the relationship between each $y_i$ and x, a simple linear regression model is developed as follows:

$$x = \beta_0 + \beta_1 y_i + \epsilon \qquad \text{S.1}$$

For each set of $y_i$, 100 pairs of $(x,y_i)$ are used in their associated linear model. The precision of the model is evaluated using the parameters of MSE (mean square error) and $R^2$ (R-squared). Their equations are as follows:

$$\mathrm{MSE} = \frac{1}{n}\sum_{i=1}^{n}(x_i - \hat{x}_i)^2 \qquad \text{S.2}$$

$$R^2 = 1 - \frac{\sum(x_i - \hat{x}_i)^2}{\sum(x_i - \bar{x})^2} \qquad \text{S.3}$$

where n denotes the number of observations in the dataset, $x_i$ denotes the value of the target variable in the i'th observation, $\hat{x_i}$ denotes the predicted value of the target variable x, and $\bar{x}$ denotes the mean value of the target variable x in the dataset.

Conceptually, MSE represents the error of the estimator or predictive model created based on the given set of observations in the sample. It measures the average squared difference between the predicted values and the actual values, quantifying the discrepancy between the model's predictions and the true observations. $R^2$ on the other hand, represents the proportion of the total variation in the data that the model can explain. For example, an R-squared value of 0.8 indicates that 80% of the variation in the dependent variable can be explained by the independent variables in the model. Therefore, a lower MSE and a higher $R^2$ value both represent a higher model accuracy, with MSE = 0 and $R^2 = 1.0$ meaning perfect model accuracy.

The 10-fold cross-validated MSE and $R^2$ values of each linear model are presented in Table S3. The model using $y_4$ as the predictor achieved the lowest MSE of 0.2345 and the highest $R^2$ of 0.9996, indicating that it provides the most accurate predictions among the single predictor models. In other words, $y_4$ (4854 nm peak) has the most linear behavior among all $y_i$'s.

Table S3. Simple 1D linear regression results with 10-fold cross-validation.

| Predictor | Cross-validated MSE | $R^2$ |
|---|---|---|
| $y_1$ | 2.4665 | 0.9962 |
| $y_2$ | 7.2260 | 0.9892 |
| $y_3$ | 1.7489 | 0.9971 |
| $y_4$ | 0.2345 | 0.9996 |

## S5. Multi-Dimensional Modeling Approaches

### S5.1 Multiple linear regression model

Now, to include all 4 peak shifts, *i.e.*, to include predicters $y_1$ through $y_4$, a multiple linear regression model is developed using all predictors as follows:

$$x = \beta_0 + \beta_1 y_1 + \beta_2 y_2 + \beta_3 y_3 + \beta_4 y_4 + \epsilon \quad \text{S.4}$$

The MSE and $R^2$ values using multiple linear regression model with 10-Fold Cross-Validation are found to be 0.0173 and 1.0000, respectively, *i.e.*, near perfect prediction model accuracy. This indicates an enhancement of over 2 orders of magnitude in the MSE value, or in the precision of estimating the bulk index, compared to the case of using $y_1$, $y_2$, or $y_3$ individually, and an enhancement of over an order of magnitude compared to the case of using $y_4$ individually in the one-dimensional linear regression modeling approach.

### S5.2 Multicollinearity assessment

Now, we assess the variance inflation factor (VIF) for all predictors, which is a statistical tool used to measure the severity of multicollinearity in a regression model [46]. This is to make sure that the accuracy of our model is not degraded because of multicollinearity among predictors. VIF quantifies how much the variance of estimated regression coefficients increases if predictors ($y_i$) are correlated (referred to as multicollinearity). Since the correlation among predictors is high (see table 1), they may have overlapped information in explaining x, therefore, we calculate VIF. VIF = 1 means no correlation, VIF = 1-10 typically means moderate correlation, and VIF > 10 means high correlation, which can be problematic in terms of reducing the model precision. A high VIF value indicates that a predictor ($y_i$) can be linearly computed using other predictors. VIF values are calculated and presented in table S4. The VIF values indicate severe multicollinearity among the predictors.

Table S4. Variance Inflation Factor (VIF) for TM polarization predictors.

| Predictor | VIF |
|---|---|
| const. | 23.1548 |
| $y_1$ | 5019.8970 |
| $y_2$ | 1722.9343 |
| $y_3$ | 1101.4025 |
| $y_4$ | 4814.3072 |

Constant is a constant value that we create as a separate column (usually 1) in VIF analysis, and its VIF is found to be 23.1584. Since this constant value inherently is not correlated with other predictors (inputs or y's) it creates flexibility and baseline for regression. Although the multiple linear regression model shows excellent predictive performance, since the calculated VIF values are high, which means a predictor can be linearly predicted from other predictors, the regression coefficients become less reliable, and hence the standard errors of the coefficients may increase. This can decrease the predictive power of the model. Therefore, we use lasso or ridge regression to reduce the possible adverse effect of multicollinearity. This is while as we observed earlier in this specific case, the presence of multiple predictors significantly increases the predictive power of predictors ($y_i$), by reducing the MSE by orders of magnitude. However, to ensure that the possible adverse effects of multicollinearity are addressed, a Ridge Regression model is employed in the following section.

**S5.3 Addressing Multicollinearity with Ridge Regression**

Given the severe multicollinearity observed among the predictors $y_i$ (table S4), Ridge Regression is employed to mitigate its effects [47]. Ridge Regression introduces an L2 regularization penalty to the loss function, which shrinks the regression coefficients and reduces the variance of the estimates. In other words, it is used to reduce overfitting by adding the penalty term to the loss function of the model.

The Ridge Regression model takes the following form:

$$\min_{\beta}\left\{\sum_{i=1}^{n}(x_i-\beta_0-\sum_{j=1}^{p}\beta_i y_{ij})^2+\alpha\sum_{j=1}^{p}\beta_j^2\right\} \quad \text{S.5}$$

where: $x_i$ is the target variable (i=1 in this case), $y_{ij}$ are the predictor variables, $\beta_0$ is the intercept, $\beta_j$'s are the coefficients, and $\alpha$ is the regularization parameter controlling the strength of the penalty. The model is trained using 10-fold cross-validation to select the optimal $\alpha$ and evaluate the model's performance. The performance of the Ridge Regression model with 10-fold cross-validation is summarized in Table S5.

Table S5. Ridge regression performance metrics for TM polarization (10-fold cross-validation).

| Metric | Mean | Standard Deviation |
| --- | --- | --- |
| MSE | 0.0173 | 0.0069 |
| RMSE | 0.1296 | 0.0226 |
| MAE | 0.1099 | 0.0197 |
| $R^2$ | 1.0000 | 0.0000 |

The Ridge Regression model achieved a very low MSE of 0.0173 (same as multiple linear regression model) with a small standard deviation, indicating consistent performance across the folds. The values of the RMSE (root mean squared error) and MAE (mean absolute error) parameters are also low, and the $R^2$ value of 1.0000 suggests that the model explains virtually all the variance in the target variable x.

**S5.4 Residual Analysis with Cross-Validated Predictions**

Residual analysis involves examining the differences between the observed values and the values predicted by the regression model, known as residuals. This analysis is essential for validating the assumptions of linear regression, including linearity, homoscedasticity (constant variance), independence, and normality of residuals [48]. By assessing these assumptions, we can evaluate the adequacy of the model and its suitability for making reliable predictions. In this section, residual analysis is performed on the Ridge Regression model using cross-validated predictions. Cross-validation provides an unbiased assessment by ensuring that each prediction is made by a model that did not use the corresponding data point during training.

A separate function is defined to perform residual analysis with the Ridge Regression model incorporating 10-fold cross-validation. The steps are as follows: 1. Training the Ridge Regression model using a range of $\alpha$ values (see equation 3) and selecting the optimal $\alpha$ via cross validation. 2. Generating cross-validated predictions using the trained model. 3. Calculating residuals as the difference between actual and predicted values. 4. Creating diagnostic plots to assess model assumptions.

Figure S1 shows the plot of Residuals vs. Predicted Values which assesses homoscedasticity and linearity. The plot shows that the residuals are randomly scattered around zero without any apparent pattern, indicating that the assumptions of linearity and homoscedasticity are satisfied.

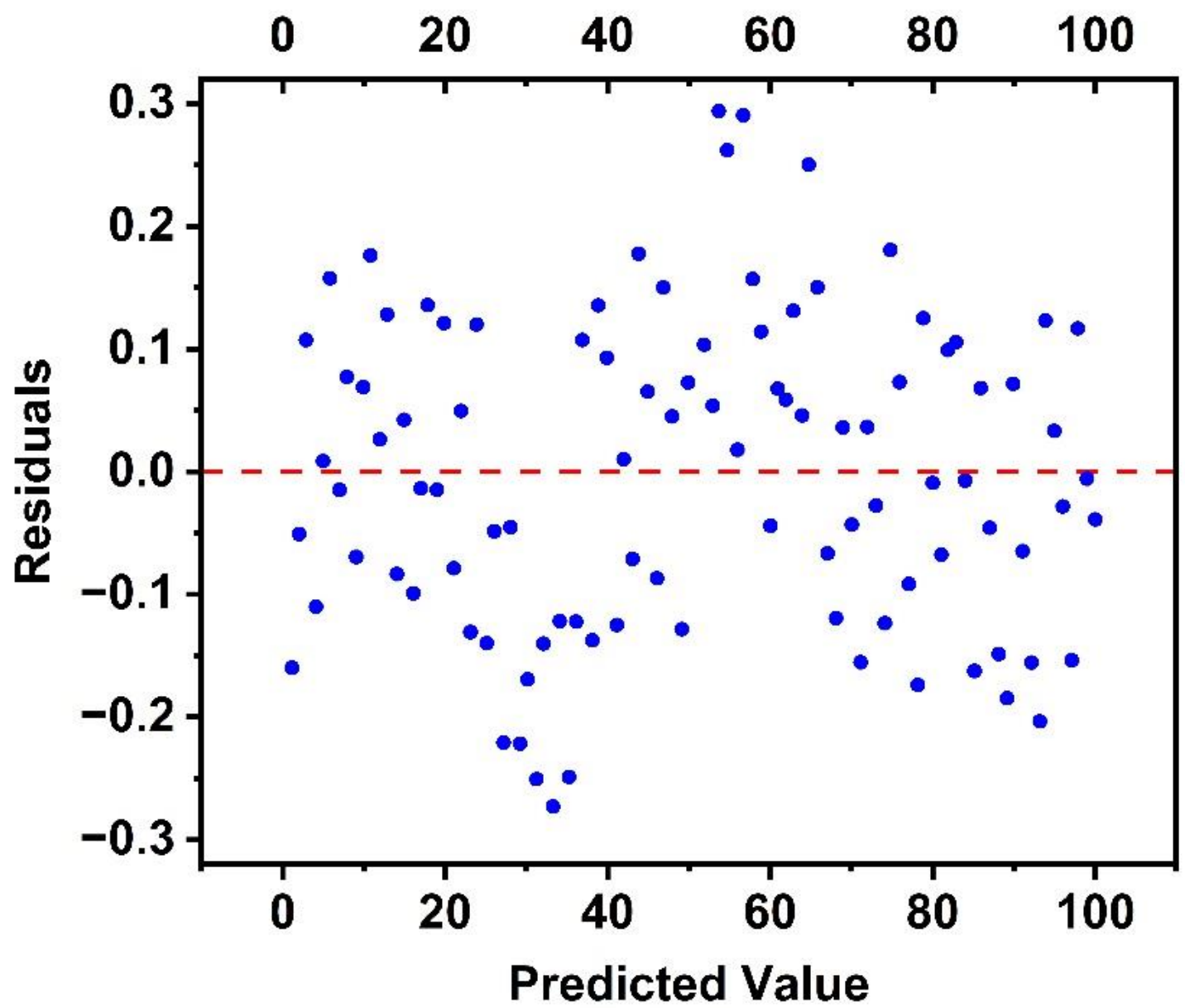

Figure S1. Residuals vs. predicted values for 4D Ridge Regression of TM polarization (10-fold cross-validated).

To complete the residual analysis, the following plots and their interpretations are presented here:

- Figure S2: Normal Q-Q Plot of Residuals, which evaluates the normality of residuals.
- Figure S3: Histogram of Residuals, to visualize the distribution of residuals.
- Figure S4: Residuals vs. Predictor $y_1$ Plot, to check for patterns indicating non-linearity or omitted variables.

As shown in Table S1, the variables have similar distributions, with means and medians closely aligned, indicating symmetric distributions. The standard deviations reflect the variability in each variable, with y2 showing the highest variability.

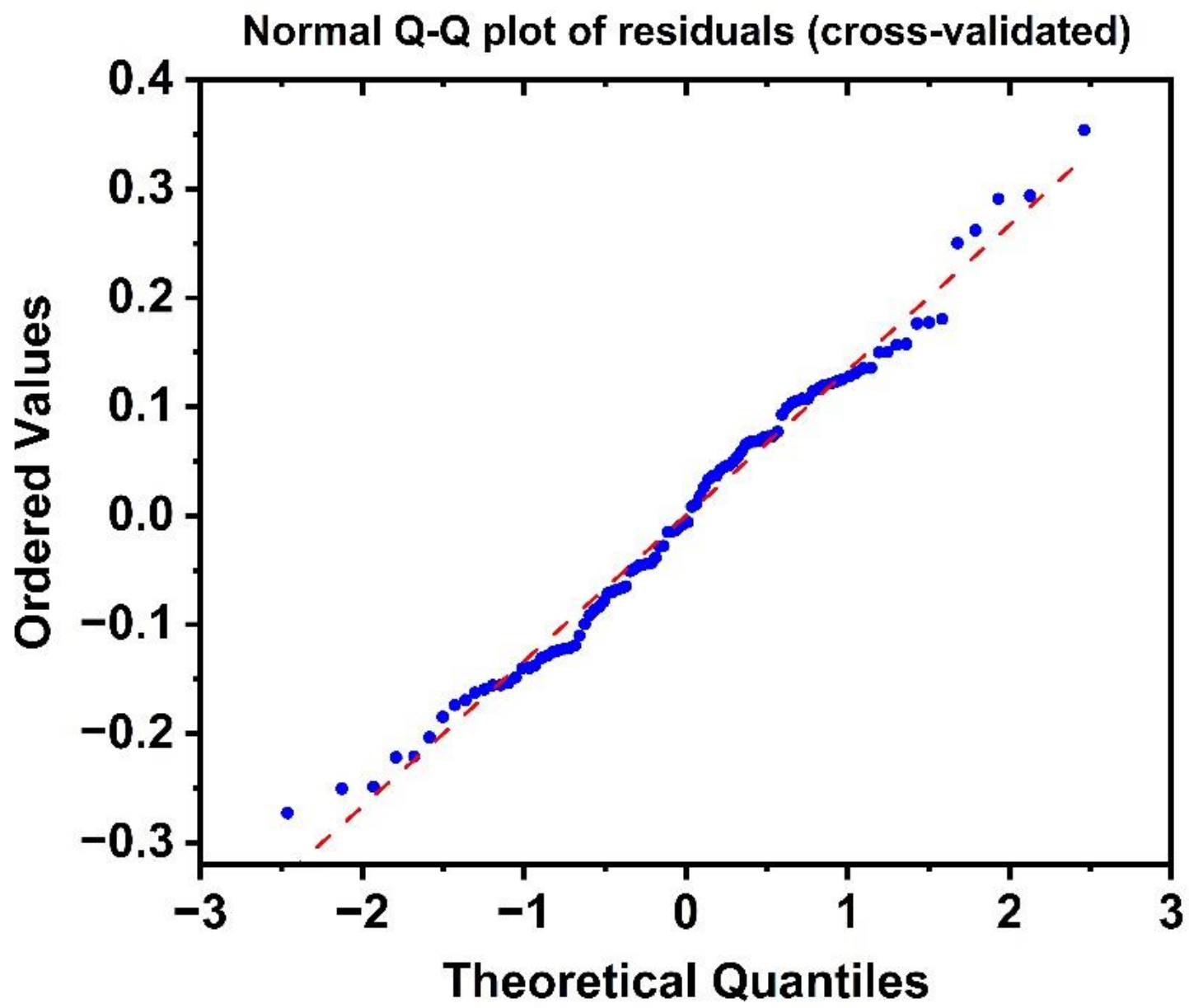

Figure S2: Normal Q-Q Plot of Residuals (Cross-Validated)

The residuals closely follow the straight line in the Q-Q plot [49], suggesting that they are approximately normally distributed, which satisfies the normality assumption.

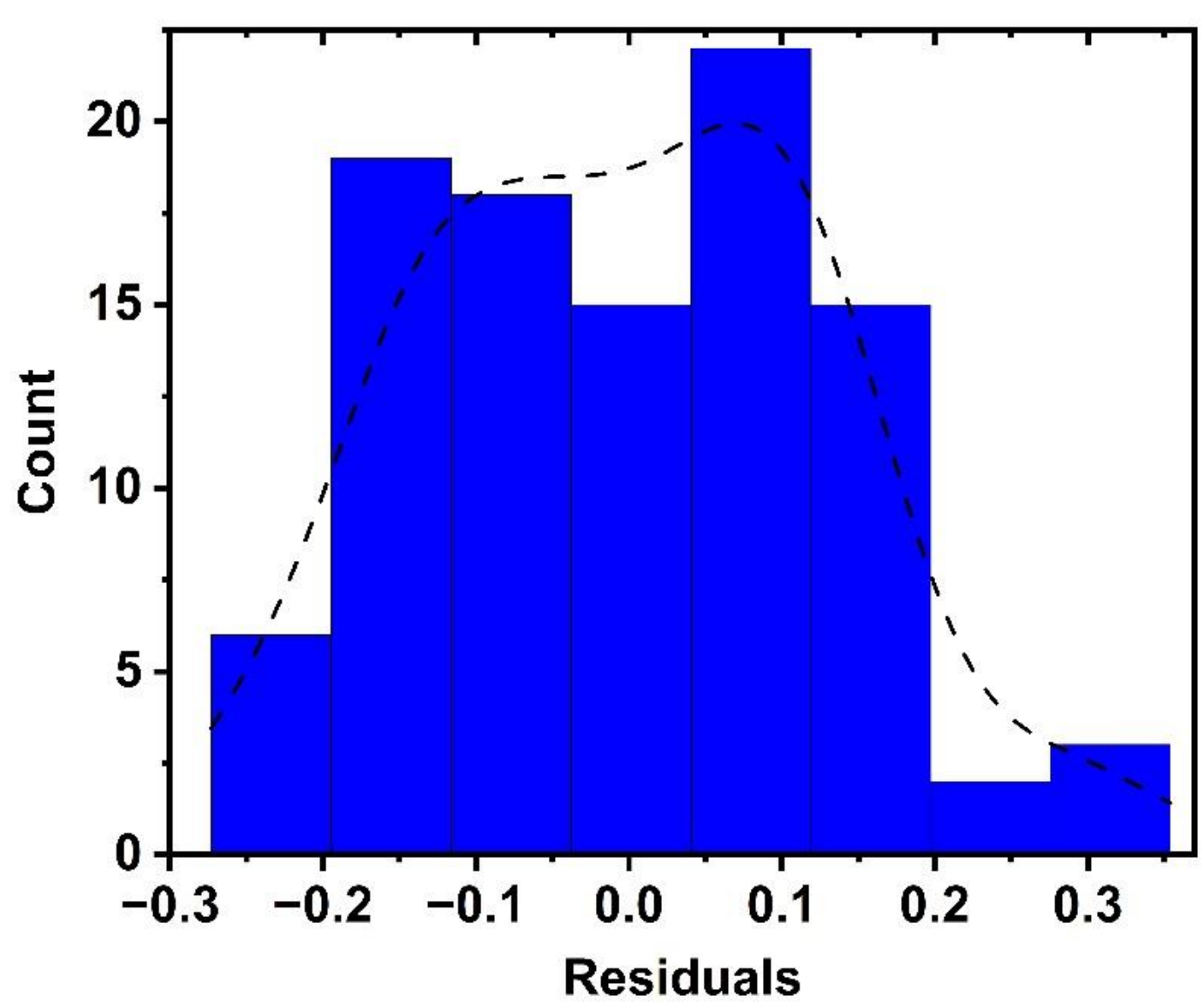

Figure S3: Histogram of Residuals (Cross-Validated)

The histogram shows a symmetric distribution centered around zero, further supporting the normality of residuals.

Residuals vs. Predictors Residuals are plotted against predictor $y_1$ to check for any patterns.

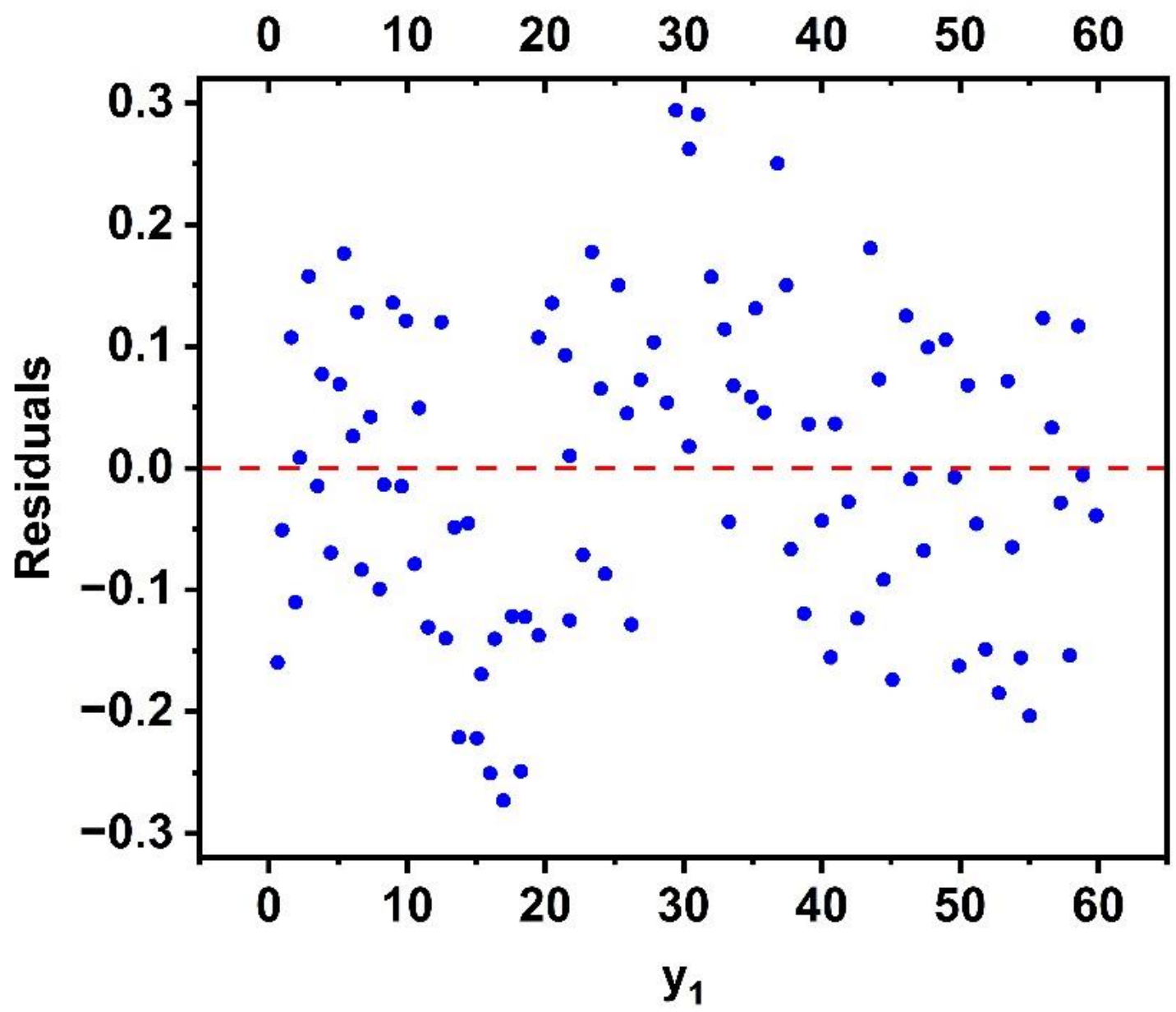

Figure S4: Residuals vs. y1 (Cross-Validated)

The residuals display no discernible patterns or trends when plotted against each predictor, indicating that the model captures the relationships adequately and that the linearity assumption holds for each predictor.

Conclusion of Residual Analysis

The residual analysis confirms that the Ridge Regression model meets the key assumptions of linear regression:

• Linearity: The relationship between the predictors and the target variable is linear.

• Homoscedasticity: The variance of residuals is constant across all levels of predicted values.

• Normality of Residuals: The residuals are approximately normally distributed.

• Independence of Residuals: Given the data collection process, residuals are assumed to be independent. These results validate the adequacy of the Ridge Regression model for predicting x using the predictors $y_1$, $y_2$, $y_3$, and $y_4$.

**S6. 4D Ridge Regression coefficients for the TM polarization**

Table S6. 4D Regression coefficients for the TM polarization

| Predictor | Coefficient |
|---|---|
| Intercept | −0.0162 |
| $y_1$ | −0.1681 |
| $y_2$ | −0.0515 |
| $y_3$ | 0.0690 |
| $y_4$ | 0.9514 |

The coefficients in table S6 indicate the expected change in x for a one-unit change in each predictor, holding other variables constant. Specifically:

• The intercept is −0.0162, representing the expected value of x when all predictors are zero.

• y1 has a coefficient of −0.1681, suggesting a slight negative relationship with x when controlling other variables.

• y2 has a coefficient of −0.0515, also indicating a slight negative relationship with x.

• y3 has a coefficient of 0.0690, indicating a small positive relationship with x.

• y4 has a coefficient of 0.9514, showing a strong positive relationship with x.

## S7. 8-dimensional Ridge Regression modeling parameters

Best alpha = 0.1748, Cross-validated MSE = 0.0090, $R^2$ = 1.0000

Table S7. Ridge Regression Performance Metrics (10-fold cross-validation)

| Metric | Mean | Standard Deviation |
|---|---|---|
| MSE | 0.0090 | 0.0041 |
| RMSE | 0.0925 | 0.0207 |
| MAE | 0.0770 | 0.0149 |
| $R^2$ | 1.0000 | 0.0000 |

Table S8. 4D Regression coefficients for TM polarization

| Polarization | Predictor | Coefficient |
|---|---|---|
| | Intercept | -0.1796 |
| TM | $y_1$ | -0.0467 |
| | $y_2$ | −0.0515 |
| | $y_3$ | 0.0532 |
| | $y_4$ | 0.5082 |
| TE | $y_{11}$ | 0.0276 |
| | $y_{22}$ | 0.0497 |
| | $y_{33}$ | 0.1859 |
| | $y_{44}$ | -0.0899 |

## S8. Feature Selection

By incorporating recursive feature elimination [50] which is a method that finds the most valuable predictors by eliminating the least valuable ones one by one, and while considering x as the target variable, and [$y_1$, $y_2$, $y_3$, $y_4$] and [$y_{11}$, $y_{22}$, $y_{33}$, $y_{44}$] as predictors for the TM and TE polarizations respectively, the feature selection results are as follows:

Top 1 predictor (meaning that if we only have the option to pick one of the predictors to predict the outcome, the following will have the most probability of predicting the outcome): y4

Top 2: $y_4$, $y_{42}$

Top 3: $y_4$, $y_{32}$, $y_{42}$

Top 4: $y_2$, $y_4$, $y_{32}$, $y_{42}$

Top 5: $y_2$, $y_4$, $y_{22}$, $y_{32}$, $y_{42}$

Top 6: $y_2$, $y_3$, $y_4$, $y_{22}$, $y_{32}$, $y_{42}$

Top 7: $y_1$, $y_2$, $y_3$, $y_4$, $y_{22}$, $y_{32}$, $y_{42}$